\documentclass[pre,amssymb,aps,twocolumn]{revtex4}
\usepackage{array}
\usepackage{epsfig}

\def\be{\begin{equation}}
\def\ee{\end{equation}}
\def\br{\begin{array}}
\def\er{\end{array}}
\def\ba{\begin{eqnarray}}
\def\ea{\end{eqnarray}}
\def\bfg{\begin{figure}}
\def\efg{\end{figure}}

\def\vr{\varrho}

\def\Bn{B^{(n)}(t)}
\def\G{\Gamma}
\def\avrho{\langle\vr\rangle}
\def\vro#1{\langle\vr_#1\rangle}
\def\d{{\rm d}}

\begin{document}
\title{Slow relaxation due to optimization and restructuring: Solution
on a hierarchical lattice}

\author{J\'anos T\"or\"ok$^{(1)}$, Supriya Krishnamurthy$^{(2,*)}$,
J\'anos Kert\'esz$^{(1)}$ and St\'ephane Roux$^{(3)}$}
\affiliation{(1): Department of Theoretical Physics,
Institute of Physics, Budapest University of Technology and Economics,
8 Budafoki \'ut, H-1111 Budapest, Hungary \\
(2): Department of Theoretical Physics, University of Oxford,
1 Keble Road, OX1 3NP, UK \\
(3): Surface du Verre et Interfaces, UMR CNRS/Saint-Gobain,
39 Quai Lucien Lefranc, 93303 Aubervilliers Cedex, France}
\thanks{present address: Santa Fe Institute, 1399 Hyde Park Road, 
Santa Fe NM 87501}

\begin{abstract}
Motivated by the large strain shear of loose granular materials we
introduced a model which consists of consecutive optimization and
restructuring steps leading to a self organization of a density field.
The extensive connections to other models of statistical phyics are
discussed. We investigate our model on a hierarchical lattice which
allows an exact asymptotic renormalization treatment. A surprisingly
close analogy is observed between the simulation results on the
regular and the hierarchical lattices. The dynamics is characterized
by the breakdown of ergodicity, by unusual system size effects in the 
development of the average density as well as by the age distribution,
the latter showing multifractal properties.
\end{abstract}

\date{\today}
\maketitle

\section{Introduction}

Slow dynamics with no separation of time scales represent a major
challenge of statistical physics. Experimental or simulation
approaches are extremely difficult, so in most cases new ideas and
models are needed for the understanding of this kind of problems.

There can be different roots of slow dynamics: Systems close to the
critical point slow down enormously due to the increasing
characteristic time. Phase separation is often accompanied by a
slow coarsening process \cite{Reichl}. In glasses the free energy
landscape is so complicated and structured that the system never finds
the global minimum and shows a history dependent behaviour called
aging \cite{glass}. Slow dynamics may also occur in intrinsically
dynamic, driven systems leading to scale free fractal structures. The
name of self-organized criticality covers a whole family of related
models \cite{bak}.

In this paper we study a model (introduced in \cite{TKKRLett}), where
the system exhibits a very slow evolution with a tendency of getting
stuck in metastable states. However the model is different from those
studied earlier in the sense that there is an element of both energy
as well as entropy barriers being present as a result of the rules of
evolution. We are able to directly link the slow evolution to a break
down of ergodicity in the dynamics. This then leads to several
interesting features of the model such as non-trivial system size
effects, a multi-fractal ``age'' distribution and a non-trivial
temporal evolution.

Motivated by our study of shearing loose granular materials
\cite{TKKRLett}, we report in this paper about a new mechanism leading
to slow dynamics. In granular materials displacement occurs in a
localized manner, in 'shear bands' which are formed along the weakest
parts of the samples. During shear, grains can rearrange themselves
and occasionally strengthen the local structure. In such a case, the
shear band finds a new configuration which avoids this zone. Based on
this picture we introduced a model where consecutive steps of
optimization (finding the weakest part of the sample) and
restructuring (random rearrangement of grains) takes place. Assuming
translational invariance in the shear direction the model becomes
two-dimensional. We have studied the model numerically in detail on
regular lattices \cite{TKKRnum}, however, it is difficult to go beyond
the simple description of numerical simulations. It turned out to show
unexpected properties including extremely slow dynamics and unusual
size dependence (breakdown of ergodicity), and it provides with
interesting predictions for the granular system.

The aim of this paper is to study this same model on a hierarchical
diamond lattice both numerically and analytically, and to compare
these results with the simulations on the euclidean lattice. We find
that despite the very different connectivities of these two lattices,
the qualitative behaviour is much the same; for some properties, there
is a quantitative matching as well. The recursive nature of the
hierarchical lattice however aids the analytical treatment greatly,
thus helping us getting a deeper understanding of the problem.

The paper is organized as follows: In the next section we define the
model in general and on the hierarchical lattice. In Section III, the
relation of the model to other problems of statistical physics is
discussed. In Section IV, the numerical results are shown and compared
to the regular lattice simulations. In Section \ref{Sec_hier} we
present the exact asymptotic solution of the model. We conclude in
section \ref{Sec_concl}. Appendix A and B contain technical details of
the calculations used in Section \ref{Sec_hier}.

\section{The model and the hierarchical lattice}

The model that we study in this paper is defined as follows: A
two-dimensional field is characterized by a single scalar parameter,
the density $\vr(x,y)$. Initially this density is generated randomly
from the distribution $p_i(\vr)$. At every step we search for the {\it
minimal path} ${\cal P}^*$ that is defined as follows: The minimal
path is a continuous, directed path $\cal P$ that spans the system in
the $x$ direction and the sum $S$ of the local densities along it, 
\be
S({\cal P})=\sum_{(x,y)\in\cal P} \vr(x,y)
\ee
is minimal among all possible paths. The minimal path is the path
${\cal P}^*$ for which $S({\cal P}^*)$ is minimum.

Once the minimal path is found the density values of the points
belonging to the minimal path $(x,y)\in{\cal P}^*$ are replaced by new
densities randomly picked from the distribution $p_r(\vr)$.

The above process is repeated as long as desired. A single time step
consists of both searching for the minimal path, as well as refreshing
the local densities along it.

In the following we restrict ourself to the case where $p_i$ and $p_r$
are uniform distributions in the interval $[0:1]$. Our model is
discretized on a lattice. In \cite{TKKRnum} we report on detailed
numerical results for various properties of the model on the Euclidean
square lattice. The analytic treatment on the square lattice has not
been possible so far. However, in this paper, we obtain exact
asymptotic solutions for the model on the hierarchical diamond
lattice.

The hierarchical diamond lattice \cite{ber} is constructed as follows:
We first consider a single bond connecting two points A and B. This
constitutes the most elementary (generation 0) lattice. The first
generation lattice is obtained by substituting the unique bond by an
elementary ``diamond'' of four bonds, i.e. two parallel connections
each consisting of two bonds in series [Fig.~\ref{Fig_genhier} b)].
The next generation is obtained recursively by the substitution of
each bond by a diamond [Fig.~\ref{Fig_genhier} c)]. Repeating the
above procedure $N$ times, produces a $N$th-generation hierarchical
lattice. This lattice has a dimension equal to 2, and hence can be
compared to its Euclidian counterpart.

All results are based on the exploitation of the recursivity of the
construction of the lattice. If one can compute the properties of an
elementary diamond and transform this into a single bond endowed with
the same, a recursive use of this procedure clearly allows the
reduction of the entire lattice back to a single bond thus determining
the global behavior. This is a real space renormalization procedure
and the structure of the lattice makes such renormalization treatments
exact. Hierarchical lattices have been widely used to study several
phenomena such as percolation \cite{hie_perco}, spin models
\cite{hie_spin}, sums of directed paths \cite{hie_dp} {\it etc}.
However, usually there is a price to pay in that the result may differ
from its Euclidean lattice counterpart. There is no general formalism
by which means to estimate the validity of hierarchical lattice
results for the Euclidean lattice. Therefore, it is necessary to
resort to numerical results to assess the similarity between the two
cases. It will be shown in the following that indeed the analogy
between the results obtained on both kinds of lattices is extremely
close. Therefore, the analytical solution obtained here provides a
better understanding of the Euclidean lattice case. 

\bfg
\centerline{\epsfig{file=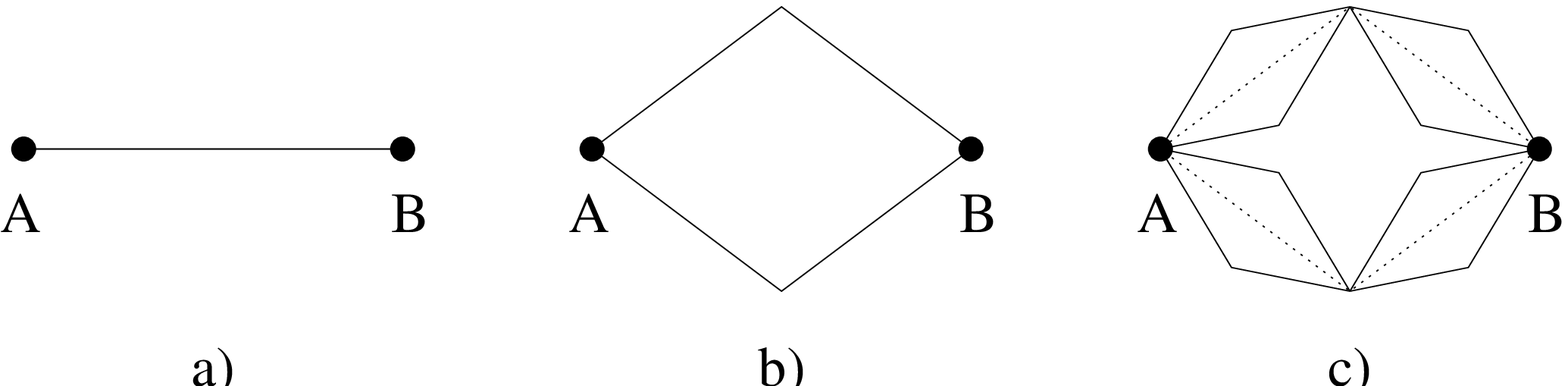,width=8truecm}}
\medskip
\caption{\label{Fig_genhier}\small
The 0th a), 1st b) and 2nd c) generations of the hierarchical lattice.}
\efg

\section{Relation to other statistical physics models}
\label{Sec_relation}

Before reporting the result of numerical simulations on the
hierarchical lattice, we point out some analogies which can be drawn
between our model and other diverse problems of statistical physics.

The rules of our model, finding the extremal directed spanning path at
every instant, is similar to finding the ground state of a directed
polymer in a random potential \cite{HalHeaRev}. However, in our case
this potential is uncorrelated only at the beginning; it changes in
time through the process already described above, of ascribing new
densities to all sites along the minimal path. The shape of the path
on Euclidean lattices is found to be self-affine in the directed
polymer problems. The model studied here changes the underlying
potential landscape in a self-organized way and naturally induces
correlations. These in turn change the self-affine exponent of the
path. This feature is studied numerically in an earlier paper
\cite{TKKRnum}.

The rules of our model can be regarded as a generalization of the
Bak-Sneppen model of evolution \cite{BakSn}, but in higher dimensions.
Indeed, the constraint of finding the minimal path and then changing
it, puts this model in a class of extremal models studied in contexts
as different as interface depinning \cite{Sneppen} and flux creep
\cite{Paczuski}. However there is an important difference between our
model and other extremal models. In the latter case, the system
(usually one-dimensional) reaches a steady state which is
`self-organized critical', in the sense that there is a power-law
distribution for avalanches in the steady state. In the case of our
model, no steady state is reached and all quantities depend on time.
As we will see, we can define avalanches which are indeed power-law
distributed, but always with time dependent prefactors. The difference
is best illustrated if we look at the simplest of these extremal
models, the Bak-Sneppen model \cite{BakSn}. This model is defined on a
one dimensional periodic array of random numbers where at every
update, the least and its neighbours are refreshed from a given time
independent probability distribution. Our model is however related to
a variant of this in which {\it only} the least is changed. In one
dimension, changing only the least does not lead to a very interesting
behaviour. However in two dimensions, as we will see, changing only
the minimal path leads to very non-trivial effects. Further, the
simple minded variation of the original Bak-Sneppen model turns out to
be very useful in solving our model on the hierarchical lattice. 

There are also connections between this model and the apparently
unrelated problem of a random walk in a disordered potential. If we
consider a one-dimensional cross-section of the model perpendicular to
the minimal path, we could imagine the point through which the
interface passes through, as the position of a random walker. The
subsequent dynamics can then be interpreted as that of a walker moving
through an initially random potential, modifying it along the way.
While the actual dynamics of the interface in two-dimensions is quite
complicated to translate in its entirety into one-dimension, it is
possible to do so in the simplest case when only corner flips are
allowed for the interface. For the walker, this simply translates to
the condition that the subsequent position of the walker is on one of
the neighbouring sites of the present one, chosen by an inequality
condition. The value at the site the walker has just left, is also
changed. We have studied such an active walker model in detail
\cite{TKKRSQD} and find that it leads to logarithmically slow
dynamics.

Because of the extremal condition used in finding the minimal path at
each time step, the solution of the model on the hierarchical lattice
uses results from extreme-value statistics \cite{Gumbel}. We also find
that the ``age'' distribution, {\it i.e.} the probability distribution
of the number of times up to time $T$, that a given site has been a
part of the shear band (and has hence been changed), has many
similarities with models of fragmentation studied in various contexts
\cite{Fracture}. 

There has been recently an upsurge of interest in systems exhibiting
an anomalously slow relaxation. Such a behavior is generically
reminiscent of a glassy behavior, and this analogy has motivated a
number of studies \cite{slow}. Just to mention one example related to
granular media, the slow compaction of sand under repeated
tapping\cite{Chicago1,Chicago2} displays analogies with glasses
obtained at different cooling rates. A number of different modelings
of this compaction process have been proposed
\cite{LudingModel,slow,model1,model2,model2a,model3,model4,%
model5,model6,model7,TetrisShake,MarioTetris,balda, robin}. Some of
these models emphasize the role of a broad distribution of energy
barriers which have to be overcome through thermal activation. This
naturally leads to the occurrence of a wide distribution of
characteristic times, with a slower and slower dynamics as the easiest
barriers are exhausted. Models of this sort have been looked at in a
wide variety of contexts ranging from trap models \cite{bouchaud} and
anomalous diffusion in the presence of quenched disorder \cite{sinai}
to constrained spin systems \cite{evans}, granular compaction
\cite{LudingModel} and ageing in soft solids \cite{cates}. Other
approaches put more emphasis on the collective nature of the necessary
rearrangement allowing for a relaxation
\cite{slow,model1,model2,model2a,model3,model4,model5,model6,%
model7,TetrisShake,MarioTetris,balda,robin}. As time passes, the
relaxation has to become more and more cooperative, and hence the
barrier is more entropic than energetic. Models with entropic
barriers have been well studied in other contexts too, such as the
Backgammon Model \cite{ritort}.

In the model we study in this paper, as we shall see, we observe a
very slow dynamics which can indeed be compared to such glassy
behavior. We do not include any temperature {\it stricto sensu},
however, the randomness of the local densities can in some way be
compared to thermal noise. The crude classification we proposed above
between energetic and entropic emphasis is not quite suited to our
model, where both aspects are simultaneously present. The necessary
cooperative nature of efficient events is included in the search for a
minimal path where all sites contribute with the same weight. However,
a {\it local} dense configuration can occur at any time, and remain
quenched thereafter for very long. This is like an energy barrier
since in order for the minimal path to go through this region, all
minimal paths with smaller energies need to be eliminated This is thus
a very rare event with the probability becoming smaller and smaller as
time passes.

To push forward the analogy with a glassy system, we will see that we
observe a breakdown of ergodicity, in the sense that the activity is
not spread uniformly throughout the system. Hence if we partition a
system into two sub parts (even for large sizes), the relative ``age''
of the two subsystems will tend to a broad distribution, and not to a
narrow one as expected for homogeneous systems. This implies that the
scaling of the compaction in both space and time is expected to be non
trivial. 

\section{Numerical results}
\label{Sec_num}

In this section we present briefly the most important numerical
results on both the hierarchical and Euclidean lattices.

The most important quantity of the system is the {\it average
density}. We define it as the mean density of all sites not belonging
to the minimal path and we denote it by $\avrho(t)$. The importance of
not including the minimal path in the average density is that this
definition ensures that $\avrho$ monotonically increases with time.
Furthermore, as we will see, at late times the minimal path mostly
remains unchanged. Since we keep refreshing the same bonds again and
again, the density along the minimal path is simply taken from the
known distribution $p_r(\vr)$ and there is no need to incorporate
this into $\avrho$.

In our case, as $p_i=p_r$ is a uniform distribution between $0$ and
$1$, it is clear from the rules that the system evolves towards the
limiting state of $\vr(x,y)=1$ everywhere. It is natural thus to plot
$1-\avrho(t)$ as done on Fig.~\ref{Fig_numavrho}.

\bfg
\epsfig{file=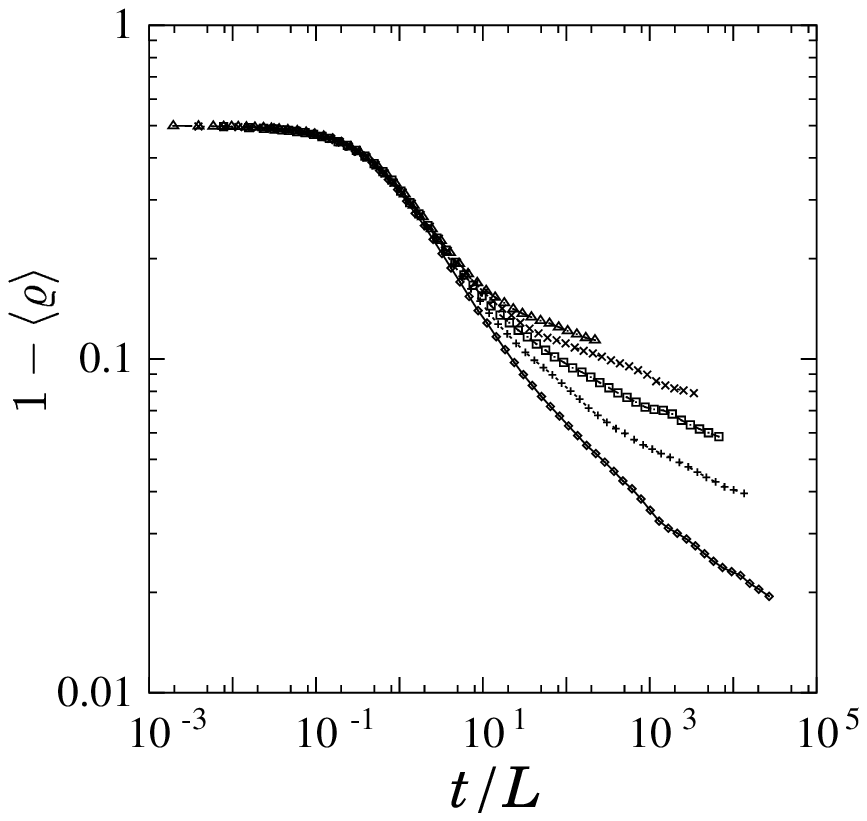,height=6truecm}
\epsfig{file=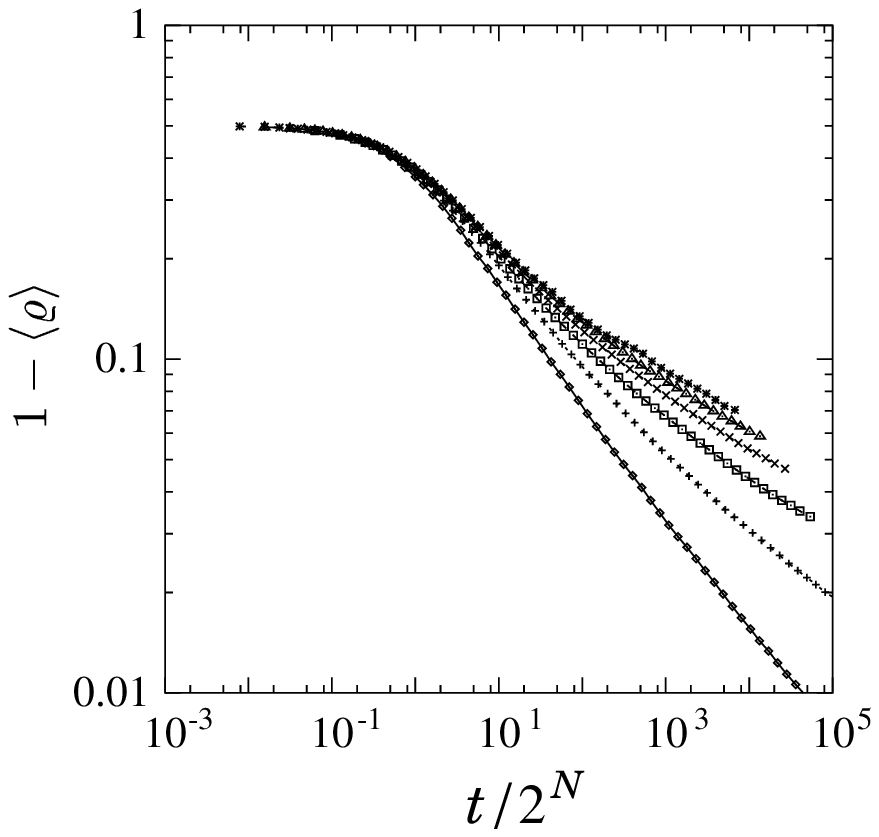,height=6truecm}
\caption{\label{Fig_numavrho}\small
The difference of the average density from its asymptotic value $1$
as a function of time. The system sizes are $L=32$, $64$, $128$,
$256$, $512$ for the square lattice (upper) and $N=2$ to $N=7$ for the
hierarchical lattice (lower) from bottom to top respectively. The
average was done over all the inactive sites in the lattice and for an
ensemble of $20$ to $1000$ samples. System size increases from
bottom to top in both cases.}
\efg

On Fig.~\ref{Fig_numavrho} we present both the $L \times L$ square
(upper) and hierarchical (lower) numerical results for $\avrho(t)$. We
can make two immediate observations: The $t/L$ ($t/2^N$) scaling works
nicely up to about unity after which a system size dependent
relaxation is observed which is slower for larger systems. The density
decay seems to be slower than any power-law. The other quantity that
we study in detail is the {\it Hamming distance}, i.e., the number of
different bonds between consecutive minimal paths. We denote this
quantity by $d$. The value of $d$ may vary from $0$ to $L$ ($2^N$).

As can be seen in Fig.~\ref{Fig_diff_t}, in both lattices, at an early
stage the mean Hamming distance is close to the system size (i.e. two
consecutive paths do not overlap at all). It then decreases
monotonically to $0$. We recall that when the distance is equal to
$0$, then the two successive conformations of the minimal path are
identical, in spite of the total renewal of random densities along
them. This indicates that minimal paths have a tendency to remain more
and more persistent as the system ``ages''.

\bfg
\epsfig{file=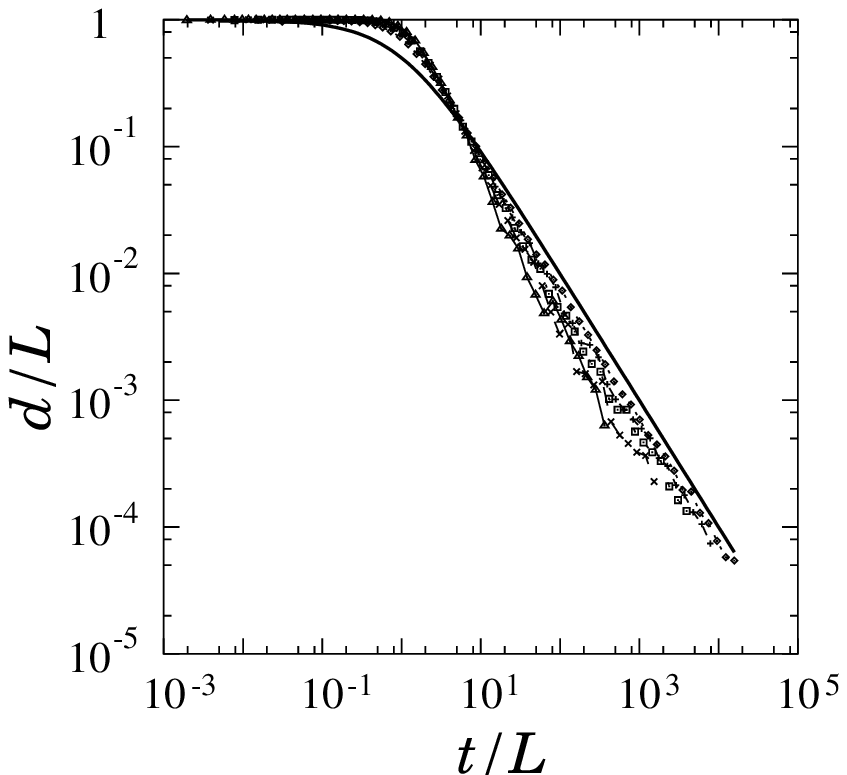,height=6truecm}
\epsfig{file=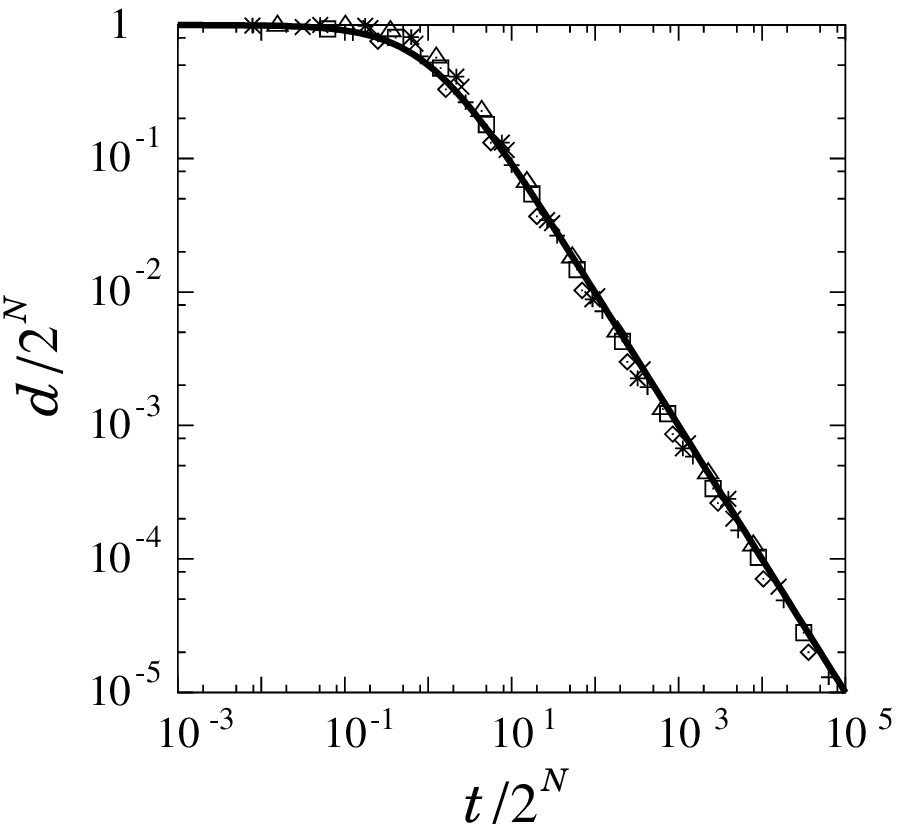,height=6truecm}
\caption{\label{Fig_diff_t}\small
The average Hamming distance versus time for square (upper) and
hierarchical lattice (lower). The same system sizes were scaled
together as on Fig.~\ref{Fig_numavrho}. On both figures, the analytical
prediction $1/(t+1)$ is plotted over the data. Note that scaling with
system size is excellent for the hierarchical lattice while it seems
to display systematic corrections for the Euclidean case.
}
\efg

\section{Analytical results on the hierarchical diamond lattice}
\label{Sec_hier}

In the following we show how some of the above listed properties of
our model can be understood analytically on the hierarchical lattice.
The techniques we use are essentially those of extreme value
statistics. 

\subsection{Summary of the solution}

We use the hierarchical nature of the diamond lattice to calculate the
average density for level $N$ knowing the results on level $N-1$.

Let us introduce the following notations: The average density of the
inactive sites in a (sub)lattice level $N$ is $\vr^{(N)}(t)$, the
density of the active sites is $x^{(N)}(t)$. In (sub)lattices we
define the active site to be the minimal path of this (sub)lattice
regardless of whether it is a part of the global minimal path or not.
We will use the indices $l$, $r$, $u$, $d$ for left, right, up, down
respectively to indicate the parts of a diamond [(sub)lattice]
corresponding to the illustration on Fig.~\ref{Fig_genhier} b,c).

A generation $N$ lattice is constructed by putting together $4$
generation $N-1$ sublattices (see Fig.~\ref{Fig_genhier}) with two
series couplings and one parallel coupling. The series coupling of
the sublattices is easily taken into account, both the density of the
minimal path as well as the density of the sites in the bulk are
simply averaged: 
\be\br{ll}
x^{(S(N-1))}(t)&\displaystyle={1\over 2}\left(x_l^{(N-1)}(t)+
x_r^{(N-1)}(t)\right)= x^{(N-1)}(t)
\\
\\
\vr^{(S(N-1))}(t)&\displaystyle={1\over 2}\left(\vr_l^{(N-1)}(t)+
\vr_r^{(N-1)}(t)\right)= \vr^{(N-1)}(t)
\er\ee
where the superscript $S$ refers to two systems coupled in series.

The next step is the coupling in parallel of two series couplets. The
density of the shear band is simply the minimum of that of the two
subsystems. The average density of the bulk contains two
contributions: the average densities of the subsystems as well as the
average density of one of the active paths (the one which is the
larger of the two contenders for the global minimal path).
\be\br{ll}\label{Eq_parallel}
x^{(N)}(t)&=\displaystyle\min\left\{x_u^{S(N-1)}(t_u),
x_d^{S(N-1)}(t_d)\right\}
\\
\\
\vr^{(N)}(t)&=\displaystyle{\left(4^{N}-2^{N+1}\right)\over 2(4^N-2^N)} 
\left\{\vr_u^{S(N-1)}(t_u)+\vr_d^{S(N-1)}(t_d)\right\}
\\
&+\displaystyle{2^N\over (4^N-2^N)} \max\left\{x_u^{S(N-1)}(t_u),
x_d^{S(N-1)}(t_d)\right\}
\er\ee
where $4^N$ is the total number of bonds, $2^N$ is the number of the
bonds in a path on a generation $N$ lattice and the multiplicative
factors in the above equation are the appropriate fractions of bonds
at generations $N$ (see Fig.~\ref{Fig_genhier}).

There is a further subtlety here. The time $t$ counts the total number
of updates at generation $N$. However the ``time'' relevant for a
subsystem at generation $N-1$ is simply the number of times the
subsystem itself has been updated. Since only one of the two systems
in parallel is updated at every instant, the ``age'' of a subsystem at
level $N-1$ is less than $t$ and is denoted by $t_u$ and $t_d$ in the
above equation. In Appendix \ref{Sec_timepart} we prove that the
relative age of either subsystems $t_u/t$ or $t_d/t$ is {\it
uniformly} distributed between 0 and 1 in the limit of a large time
$t$. Thus we rewrite the second equation of (\ref{Eq_parallel})
\be\br{l}
\label{Eq_parmean}
\vr^{(N)}(t)\displaystyle={\left(4^N-2^{N+1}\right)\over (4^N-2^N)}
{1\over t}\int_0^t \vr^{(N-1)}(t') \d t'+ \\
\displaystyle+{2^N\over (4^N-2^N)}{1\over t}\int_0^t 
\max\left(x^{(N-1)}(t'),x^{(N-1)}(t-t')\right) \d t' .
\er\ee

The second term in Eq.~\ref{Eq_parmean} comes from the competition
between the minimal paths in the two subsystems coupled in parallel.
Only one of these is the global minimum and the larger has hence to be
incorporated into the density of the system. At every timestep that a
subsystem is updated the minimal path of that subsystem can switch to
either side. Since the two parallel subsystems are entirely disjunct
the path changes sides if the mean of the random numbers generated
along the minimal path is larger than the minimal path in the other
subsystem. This competition is present at all levels of the
hierarchy.

The problem of the minimal path in the parallel coupling can be thus
described by a simple model that we call the {\it two site model}.
The two site model is defined as follows: There are two sites, each
with a single value generated by a random number drawn from a given
distribution $p(x)$ (in our case at level $N$ it is the sum of $2^N$
independent random numbers each of which is taken from the uniform
distribution between $0$ and $1$). We choose the site with the smaller
value and refresh it with a random number generated from the same
distribution \cite{zia}. The dynamics consists of repeating this
procedure. Important features of this problem turn out not to depend
on the distribution $p(x)$, since the entire evolution is only based
on the ordering of the values. As a result one can map any bounded
distribution onto a uniform one and preserve the same history of the
activity. It is thus easy to deduce that the probability of having an
active site in one subsystem for a given time, knowing the age of the
system, is independent of $p$. We present in Appendix
\ref{Sec_twositen} an analytical derivation of relevant properties of
this problem.

\subsection{Age distribution}

We have seen that as a result of the parallel coupling, the time spent
in one subsystem, or the ``age'' of a subsystem, $\theta$, differs
from the actual time $t$ and that $\theta/t$ is uniformly distributed
between 0 and 1 in the two-site model. Repeating the above argument
from the entire system down to a single bond, we can extract the
statistical distribution of ages relative to the total time.

Let $p_N(T;t)$ be the statistical distribution that a given bond was
updated {\it exactly} $T$ times at time $t$ in a lattice of generation
$N$.  Using the above argument, we can relate these distributions of
different generations through the relation:
\be
p_{N+1}(T;t)= \int_T^t {p_N(t';t)\over t'} \d t'
\ee
with $p_0(T;t)=\delta(t-T)$ and $N \geq 1$. In other words the age of
the bonds or subsystems can be obtained by a ``fragmentation''
process: At every level to get the age of the upper and lower arms in
the parallel coupling the age of the diamond is cut into two pieces
with a uniform distribution. Not surprisingly equations similar to the
above are well known in the context of models of fragmentation
\cite{Fracture}. 

The solution of the above recursion is :
\be
p_N(T;t)={1\over t}{\left(\log(t/T)\right)^{N-1}\over (N-1)!}
\ee
for $T\le t$. Introducing the relative age $\theta=T/t$, we observe
that the above distribution becomes independent of the time $t$ (the
$1/t$ prefactor is absorbed in the measure $d\theta=dT/t$).

We note here that models of fragmentation which are described by
similar equations usually look for a steady state solution {\it i.e.}
an $N$ independent solution at late times. However, in our case, as
explained below, the $N$ dependence is crucial and has necessarily to
be kept. Further the order in which $N$ and $t$ are taken to infinity
is very important as well.

It is interesting to note that the above distribution can be simply
expressed in the framework of multifractality, which was introduced to
characterize the scale dependence of statistical distributions. This
analysis naturally provides a generalized ``dimensional analysis'' of
a local quantity $x$, with a distribution $p_L(x)$. We introduce the
scaling index $\alpha$ and associated fractal dimension $f(\alpha)$ of
the support of the set of $x$ values defined through
\be\left\{\br{ll}
x &\propto L^\alpha\\
xp_L(x)&\propto L^{f(\alpha)-d}
\er\right.
\ee
where $d$ is the space dimension. Alternatively
$\alpha=\log(x)/\log(L)$ and $f(\alpha)=d+\log(xp_L(x))/\log(L)$. In
our case, the local quantity $x$ is the relative age
$\theta=T/t$ and $d=2$, thus
\be\br{lcl}
\alpha &=&\displaystyle {\log(\theta)\over \log(L)}\\
f(\alpha) &=&\displaystyle 2+\alpha+\left({1\over\log(2)}-
{1\over\log(L)}\right)
\bigg\{\log(-\alpha)+\\
\\
&&\displaystyle\left.+1+\log(\log(2))-{\log(\log(L/2))\over\log(L)}\right\}
\er\ee
where we have used the Stirling formula, assuming $2^N=L\gg 1$. In this
limit, we have
\be\br{lcl}
f(\alpha)&=&\displaystyle \alpha+{\log(-\alpha)\over\log(2)}
+{1+2\log(2)+\log(\log(2))\over\log(2)}+\\
\\
&&\displaystyle +{\cal O}\left({1\over\log(L)}\right)
\er\ee
where in the limit of an infinite system size, $L\to\infty$, the
correction term ${\cal O}(1/\log(L))$ vanishes. Due to this formalism
we arrive at a {\it system size independent description of the
distribution of relative age}, although the distribution itself
depends on $L$. Moreover, the interpretation of the formalism is
rather natural. The subset of sites whose age scales as a power-law of
the system size $\theta\propto L^{\alpha}$ has a fractal dimension
$f(\alpha)$.

\bfg
\centerline{\epsfig{file=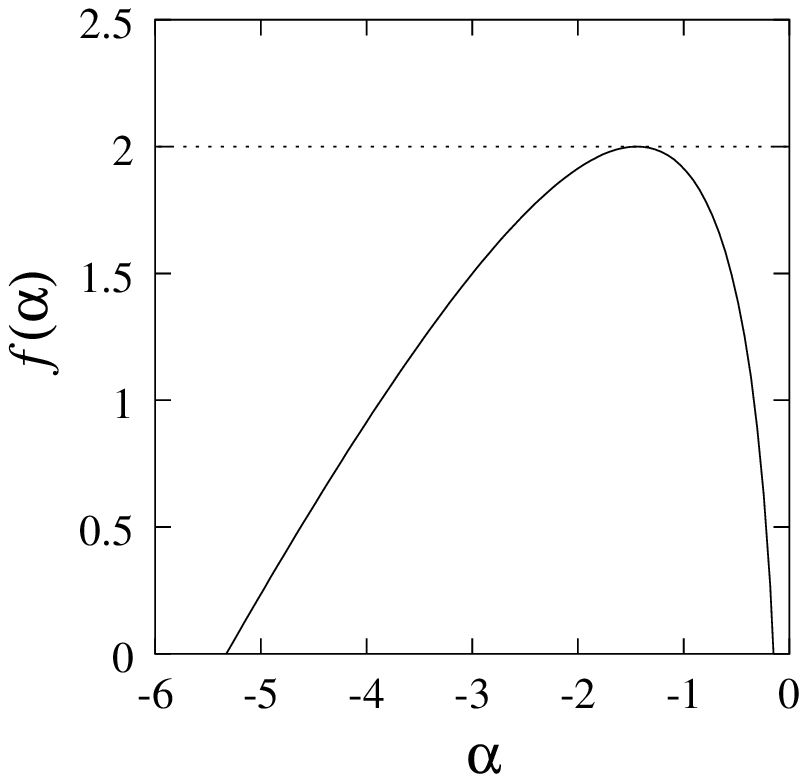,height=6truecm}}
\medskip
\caption{\label{Fig_multif}\small
Multifractal spectrum of the (relative) age distribution in the 
hierarchical lattice. $\alpha$ gives the scaling exponent of the age
with the system size, and $f$ the corresponding fractal dimension of the
support of the set of sites contributing to a given $\alpha$.
}\efg

Figure \ref{Fig_multif} shows the asymptotic form of the multifractal
spectrum. The range of $\alpha$ values corresponding to a positive
fractal dimension is $\alpha_{min}\approx-5.33$ and
$\alpha_{max}\approx-0.15$. The scaling exponent characterizing the
maximum number of sites $\alpha_0$ is the one for which $f$ is
maximum, i.e. $f=2$, and hence $\alpha_0=-1/\log(2)\approx-1.44$. Let
us emphasize that this description is only valid for very large times.
Otherwise, the finite cut-off in the time distribution will affect the
multifractal spectrum. Moreover, we have discarded correction terms
which will disappear as $1/\log(L)$, i.e. very slowly. This may render
this spectrum difficult to observe numerically.

This analysis shows that the relative age $\theta$ does not scale with
$L$ in a unique fashion. When computing a moment of order $m$, only
one scaling set dominates. The precise value of this dominant $\alpha$
depends on $m$. It corresponds to the condition $\d f(\alpha)/\d
\alpha=-m$ or
\be
\alpha(m)={-1\over(m+1)\log(2)}
\ee
unless the corresponding value of $f$ is negative. The moment then
varies as
\be
\langle \theta^m\rangle \propto L^{\tau(m)}
\ee
where 
\be
\tau(m)=f(\alpha(m))-d+m\alpha(m)
\ee
In our example, 
\be
\tau(m)=-{\log(m+1)\over\log(2)}.
\ee

\subsection{Average density}
\label{Sec_AvDens}

Now we can use the above results to get the final form of the time
evolution of the average density.

The recursion relation in Eq.~\ref{Eq_parmean} is composed of two
terms. Let us study the first term. As the integration operation is
additive we can consider separately all the components of a lattice of
generation $N$; from subsystems of generation $N-1$ right upto
individual bonds.

First we consider the generation $1$ lattice, the simple diamond. The two
site model gives the exact time dependence of the inactive bonds in
the asymptotic limit (see Appendix \ref{Sec_twositen}) which is
$1-B^{(2)}(t)\propto 1/\sqrt{t}$. In order to get the contribution of
these bonds we have to complete the integral of Eq.~\ref{Eq_parmean},
the expectation value of $T^{-1/2}$, with the correct ``age''
distribution of these subsystems. Note that we calculate the age of a
diamond (level $1$ object) not a bond. Therefore in a level $N$ system
we shall use $p_{N-1}(T;t)$ for the age distribution.
\be\br{ll}
\label{eq_intmulti}
1-\vro{1}&\displaystyle\propto t^{-1/2} \int_{0}^1
\theta^{1/2}p_{N-1}(T;t)~\d\log(\theta)\\[10pt]
&\displaystyle= t^{-1/2} \int_{0}^1
\theta^{1/2}{(-\log(\theta))^{N-2}\over(N-2)!}\d\log(\theta)\\[10pt]
&\displaystyle=t^{-1/2}~2^{N-1}
\int_{0}^\infty {x^{N-2}\over(N-2)!}\exp(-x)\d x\\[10pt]
&\displaystyle=t^{-1/2}L/2 
\er\ee
where $\vro{i}$ is the contribution of sublattices of generation $i$ to
the average density.

The above result has two important implications. First, the time
dependence of the sublattice, $1/\sqrt{t}$, is preserved on the global
scale. Secondly, the statistical distribution of ages gives rise to a
system size dependence, {\it i.e.} a power-law of $L$, which in the
above example displays a trivial exponent $1$. More generally, this
exponent is $\tau(-1/2)$ as derived above.

The above expression accounts for about half of the bonds. The next
term which enters in the coupling is the inactive minimal paths in the
level $2$ subsystems (Fig.~\ref{Fig_genhier} c). The length of these
paths is $2^2=4$ bonds. The two-site model predicts an asymptotic
$t^{-1/4}$ time dependence for $1-B^{(4)}(t)$. Thus here we have to
use the moment of order $-1/4$ and the age at generation $N-2$:
\be\br{ll}
1-\vro{2}&\displaystyle\propto t^{-1/4} \int_{0}^1
\theta^{1/4}p_{N-2}(T;t)~\d\log(\theta)\\[10pt]
&\displaystyle=t^{-1/4}~(4/3)^{N-2}
\int_{0}^\infty {x^{N-3}\over(N-3)!}\exp(-x)\d x\\[10pt]
&\displaystyle=t^{-1/4}(3/4)^2~L^{\log(4/3)/\log(2)}
\er\ee

We see in this example a non-trivial scaling with the system size and
a slower time dependece.

We can carry out this same procedure for higher genration of
subsystems. The length of the path in a level $i$ sublattice is
$n=2^i$ thus their contribution is:
\be
\label{eq_multires}
1-\vro{i}\propto t^{-1/n}(1-1/n)^i L^{-\log(1-1/n)/\log(2)}
\ee

Thus we observe that the scaling of the mean density can be cast into
the form of a sum of power-laws with a vanishing exponent $1/L=2^{-N}$
and thus a slower and slower decay to zero. Each of these terms has a
prefactor which exhibits a different scaling with $L$, and hence, the
aging of different system sizes cannot be accounted for by a simple
reduced time such as $t/L$. The latter only holds for the first subset
(half of the system size), and not the successive hierarchy of minimal
paths. This argument explains why the time evolution of the mean
density seemed to follow a unique curve when plotted as a function of
$t/L$ for early times. However, as time increases, we note a breakdown
of this simple scaling, and larger systems shows a slower and slower
increase in the average density. It is interesting to note however
that for (extremely) large times, (i.e. vanishing exponent $1/n$), the
moment will depend only on the combination $t/L^{1/\log(2)}$. This
exponent which appears in $L$ is the $\alpha$ value of the largest
fractal dimension, $f=2$, $\alpha=\alpha_0$ in the multifractal
spectrum.

The above analysis is, however, valid only for very late times, after
a long transient. The sum of $m$ identically distributed random
variables (when each individual variable is taken from the uniform
distribution) can be approximated by a power-law distribution, only at
late times. Before that, it is well approximated by a Gaussian, by the
central limit theorem. It is only when we are pushed to the tails of
the distribution that the power-law regime occurs. However the
extremising rule makes this inevitable, though after a long transient.
For instance for a generation 3 minimal path, consisting of $8$ bonds
in series, this transient ageing period lasts for about $t\sim
8!\approx 40000$ time steps. $t$ thus has to be much larger than this
so that the lower limit of the integral in Eq.~\ref{eq_intmulti} can
be taken to $0$. We see that our computation becomes strictly
applicable only for extremely late times. 

Finally we put together all the information we have on the increase of
the density of the inactive bonds of the hierarchical lattice {\it
i.e.}:

\begin{enumerate}
\item The lattice is a collection of two site systems from level 
$1 \le i \le N $ where the ``ageing'' of a bond in any level is
given by Eq.~\ref{Eq_twositeresult};
\item The number of bonds in each level gives a prefactor of
$2^{N-i}/(2^N-1)$.
\item The age distribution at each level results in an additional factor
as in Eq.~\ref{eq_multires}.
\end{enumerate}

We finally get the following result for the average density:

\be\label{Eq_hierfullres}
\vr^{(N)}(t)=1-\sum_{i=1}^N
\left({2^{N-i}\over 2^N-1}\right)
{a(i) L^{z(i)}\over(t)^{b(i)}}
\ee
where (using $n=2^i$ and $L=2^N$ for a lattice of level $N$):

\be\br{rl}
a(i)=&\displaystyle
\G(1/n)(n!)^{1/n} n^{-2} (1-1/n)^i
\\
b(i)=&1/n\\
z(i)=&\displaystyle -{\log(1-1/n)\over\log(2)}
\er\ee

Figure \ref{Fig_hier} visualizes this result compared to the numerical
data.

\bfg
\centerline{\epsfig{file=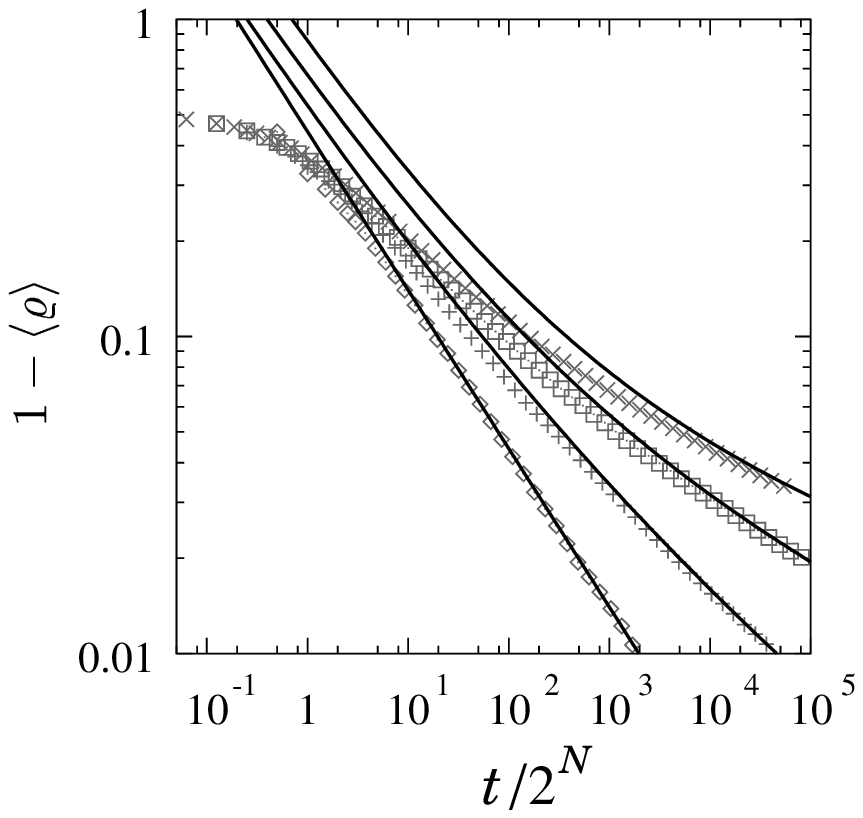,height=6truecm}}
\medskip
\caption{\label{Fig_hier}\small
The test of the analytical result. The numerical data are plotted with
symbols the corresponding analytical with solid lines. The
hierarchical level is $N=1$, $2$, $3$, $4$ from bottom to top
respectively.
}\efg

\subsection{Intermediate time behavior of the average density}

For a minimal path of length $\ell=2^n$, we are interested in the
maximum value of the sum of $\ell$ random numbers over a number of
realizations equal to the age $t$. For large $\ell$, the distribution
of the average element in the sum converges towards a Gaussian of
average $1/2$ and standard deviation $1/\sqrt{12\ell}$. The
expectation value of the largest such element over a time $t$ is thus
such that
\be \int_S^\infty {2\sqrt{3\ell}\over\sqrt{2\pi}}
\exp\left({-6\ell(x-1/2)^2}\right) \d x\approx{1\over t}
\ee
It is important to note that this expression is valid for large $\ell$
and moderate $t$, whereas we previously considered the limit of large
$t$ and moderate $L$. The order of the limits plays a crucial role.
The mean value of the densities along the minimal path thus departs
only very slowly from $1/2$. This slow change of the density of the
sites along the minimal path in turn plays a crucial role in the very
slow decay of the mean density in spite of the vanishing fraction of
bonds involved. The departure from $1/2$ varies roughly as
$\sqrt{\log(t)}$. Taking such a form into account, we see that the
average density does not converge to 1 any longer, just as if some
bonds were quenched close to their average value $1/2$, up to a very
slowly evolving correction. Thus numerically, one can achieve a
reasonable fit of the evolution of $\vr$ to values different from 1.
However, as the time window is enlarged, the effective asymptotic
$\vr$ increases. Reciprocally, extending the system size, this
asymptote decreases. Thus, in spite of the quality of the fits which
can be produced this way, we underline the fact that such an approach
is only applicable to a fixed time or system size window.

\subsection{Hamming distance}
\label{Sec_hierHam}

Let us now consider the overlap function shown in
Fig.~\ref{Fig_diff_t}. For the hierarchical lattice, the overlap has a
simple interpretation. We have seen that, at least for large times,
most of the activity essentially takes place along the same path.
However, from time to time, the minimal path jumps from one
conformation to another, whose distance to the previous one is
quantified by the Hamming distance. The scarcity of the jumps allows
us to neglect the occurrence of simultaneous multiple jumps. Let us
define the probability, $P_L(d,t)$, as the probability that a jump
equal to $d=2^n$ takes place, {\it i.e.} the probability that the
current path differs from the previous one by $d$ sites in a system of
size $L=2^N$. This quantity, apart from containing information about
how the average value of $d$ changes with time, is also the natural
analogue of an ``avalanche distribution'' in this model. As will be
seen further down, this quantity does indeed decay for large times as
a power-law of the distance $d$ like in many other self-organized
critical models. However the distribution has a time-dependent
prefactor unlike other models with a true steady state.

For $n=N$, we have to consider a jump at the largest scale available
in the system. At this level, the lattice can be coarse-grained as a
generation 1 lattice. The probability for such a jump to occur is
equal to the probability that in a two-site model, the activity moves
from one site to the other one. We show in the appendix that this
probability, is equal to $1/(t+1)$, and thus for large times,
\be
P_L(d=2^N,t)\approx 1/t
\ee

Let us now consider a smaller jump size, i.e. $n=N-1$. This means that
one half of the actual minimal path should move to a different
configuration. Thus we focus on a subsystem of size $L/2$, whose age
is $T$. In the appendix, we show that the probability for such an age
is $q(T,t)=2(T+1)/(t+1)(t+2)\approx 2T/t^2$, (see
Eq.~\ref{eq_twositjump}). Moreover, we have two such subsystems in
series and thus the probability that the Hamming distance is $L/2$ in
a system of size $L$ is approximately twice the probability that in
one subsystem the Hamming distance is equal to the system size,
$P_{L/2}(L/2,t)$. Integrating over all times $T$ with the above
probability we have
\be
P_L(L/2,t)\approx 2\int_0^t {2T\over t^2} {1\over T}~\d T=4/t
\ee

\bfg
\centerline{\epsfig{file=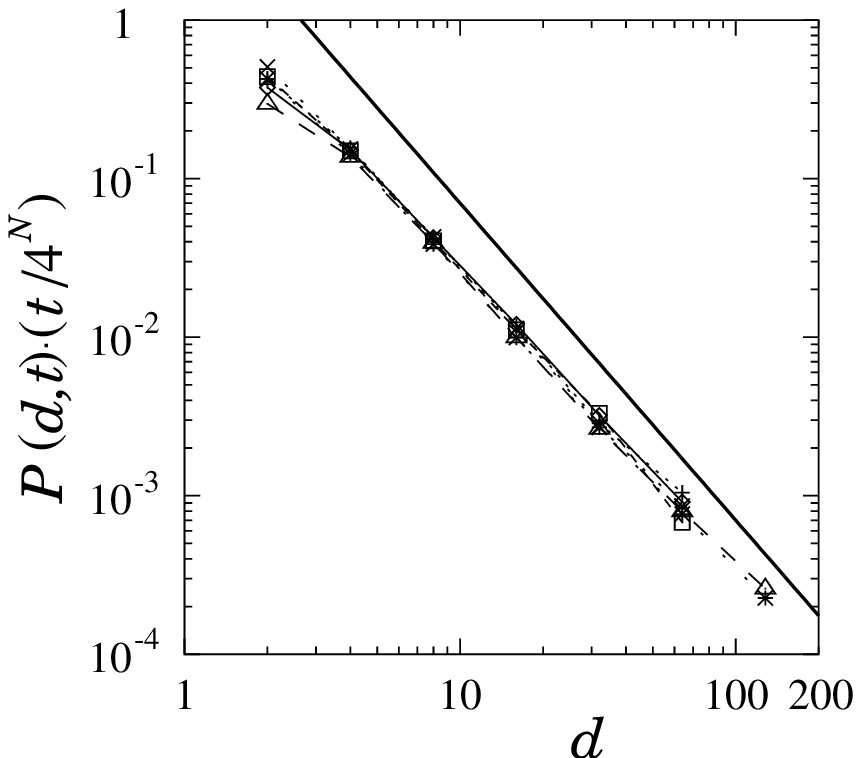,height=6truecm}}
\medskip
\caption{\label{Fig_hiesc}\small
$P(d,t)t$ scaled by the system size for the hierarchical lattice of
generation $N=6$ at times $t/L= 100$, $200$, $300$ and $1000$ and for
generation $N=7$ for times $t/L=100$ and $500$. Time is measured in
terms of the system size and the straight line has a slope $-2$. 
}
\efg

For smaller jumps, we can repeat the same argument recursively, to
obtain

\be
\br{rl}
P_L(2^n,t)\approx&\displaystyle
2^{N-n} \int_0^t\int_0^{T_1} \cdots \int_0^{T_{N-n+1}}
{2T_1\over t^2} {2T_2\over T_1^2} \cdots \times \\
\\
&\displaystyle\times\cdots {2T_{N-n}\over T_{N-n+1}^2}
~\d T_1 \cdots\d T_{N-n}=\\
\\
=&\displaystyle{2^{2(N-n)}\over t}= {L^2\over{t d^2}} 
\er\ee

As one can see this result agrees well with the numerical results
shown on Fig.~\ref{Fig_hiesc}. Moreover, this expression is to be
compared with the Hamming distribution obtained with a logarithmic
measure for $d$ on the Euclidean lattice \cite{TKKRnum}, which has the
same functional form. Hence this distribution is the same for the two
lattices despite their connectivities being very different. Note also
that this quantity scales perfectly with the system size and is
described at all times by the reduced time distribution $t/L$ as
evidenced in Fig.~\ref{Fig_diff_t}. Thus the behaviour of the Hamming
distance is much simpler than the slow density increase in the system.

\section{Discussion and conclusion}
\label{Sec_concl}

We have presented simulation results and an asymptotic analysis of the
behavior of the the optimization and restructuring model on the
hierarchical lattice. The two lattices are very different in
structure, yet they also exhibit remarkably similar features.

The Hierarchical lattice is easier to analyse due to its recursive
structure. For example, the very specific connectivity of sites on
this lattice ensures that large jumps are always possible, though rare
at late times. In the Euclidean lattice, these are strongly
suppressed by a further feedback effect: the localization of the path
limits the density increasing effect of the dynamics to a small region
around the path which in turn intensifies the localization. This
feature also results in the density map being very different in the
two cases. In the Euclidean case \cite{TKKRnum} the inhomogeneities in
a late-time snap shot of the system are much more enhanced. Another
difference is that changes in conformation in the hierarchical lattice
are organized in a strictly hierarchical way. This is clearly not so
on the square lattice, where randomness and self-organization play an
important role. Nevertheless, the overall behavior of the two type of
lattices is remarkably similar.

These similarities are most apparent in the time and size dependence
of the average density (Fig.~\ref{Fig_numavrho}). In both cases we
have a data collapse for short times, while for longer times the
dynamics becomes slower and slower as the size of the system
increases. For the hierarchical lattice, we have obtained an analytic
expression for the scaling of the Hamming distance, for the local age
distribution, and its multifractal spectrum, and the asymptotic
average density evolution with time. In particular, the mechanism
behind the breakdown of ergodicity, and the unusual size dependence of
the density evolution can be traced back to the multifractal
distribution of age. The latter provides a novel scenario for
``glassy'' ageing.

\section{Acknowledgments}

This work was supported by EPSRC, UK, OTKA T029985, T035028. SK would
like to thank Robin Stinchcombe for useful discussions.

\appendix

\section{Time partitioning in the two-site problem}
\label{Sec_timepart}

We consider here the two-site model and prove that after the elapse of
a time $t$, the probability that a given site has been visited $T$
times is uniformly distributed between $0$ and $t$. The following
proof is valid for {\em any} distribution $p(x)$. Let us compute
$q(T,t)$, the probability that {\it the active site} has been
refreshed $T$ times up to time $t$. Let the value of the recently
refreshed site be denoted by $x$ and the inactive site at $t$ be
$b_t$. At time $t$ two things may happen: 
\begin{itemize}
\item[i)] either $x<b_{t-1}$, and thus $T_{t}=T_{t-1}+1$. This happens
with probability \hfill\break $(t+1)/(t+2)$
\item[ii)] or $x>b_{t-1}$, and thus $T_{t}=t-T_{t-1}$ with probability
$1/(t+2)$.
\end{itemize}
Note that for any distribution $p(x)$ the activity change can only be
due to the fact that the largest generated random number up to time
$t$ is at instant $t$. This happens with probability $1/(t+2)$ because
in an independent time series of random numbers the largest number is
equally likely to be anywhere. At time $t=0$ we have to initialize
the system by generating two random numbers for the two sites. This is
the reason for the shift in time from $t$ to $t+2$.

Now we can write down a simple evolution equation for $q(T,t)$
\be
q(T,t)=q(T-1,t-1){t+1\over t+2}+q(t-T-1,t-1){1\over t+2}
\ee
The general solution of the recursion can be written as
\be
q(T,t)=A{(T+1)\over(t+1)(t+2)}+B
\ee
It is simple to compute $q(t,t)$ from the above recurrence, and get 
\be
q(t,t)={2\over(t+2)}
\ee
and thus $B=0$ and $A=2$. Thus finally
\be
\label{eq_twositjump}
q(T,t)={2(T+1)\over(t+1)(t+2)}
\ee

Now, the number of updates of the other site is simply $q(t-T,t)$,
thus we can formulate the probability distribution that a site has
been updated $T$ times:
\ba
r(T,t)&=&{1\over 2}\left(q(T,t)+q(t-T,t)\right)=\nonumber\\
&=&{(T+1)+(t-T+1)\over(t+1)(t+2)}={1\over(t+1)}
\ea
which is independent of $T$. Thus $r(T,t)$ is uniform.


\section{Average density of the inactive site in the two site model}
\label{Sec_twositen}

Since the hierarchical lattice can be considered as a set of two-site
systems, we need to consider the distribution of the maximum in a two
site model in the case when each of the sites is taken from a
distribution $p_n(x)$. Here the subscript $n$ denotes that this is the
distribution for a sum of $n$ independent random numbers each of which
is taken from the uniform distribution. We need only consider the
case when $n=2^N$ as the length of the hierarchical lattice can only
be of this form. 

Unfortunately, $p_n(x)$ is difficult to formulate in a general way.
For large $n$ this is a Gaussian for moderate values of $x$. However
very close to the extremes $0$ and $1$, it is a power-law as we will
see below. It is this regime which is asymptotically reached and hence
relevant for our purposes. We hence only consider the regimes $x<1/n$
or $x>1-1/n$ (cases when only one number out of $n$ may reach its
extreme value $1$ in the $x<1/n$ case and $0$ in the $x>1-1/n$ case).

Let us recall the formula for the average value of the largest
generated number up to time $t$ when each of the individual numbers
$x$ is taken from a distribution $p_n(x)$:
\be
\Bn=(t+1)\int_0^1xp_n(x)P_n^t(x)\d x,
\ee
where $P_n(x)$ is the cumulative distribution of $p_n(x)$. Thus
$P_n^t(x)$ accounts for the probability that the $t$ other numbers are
less than $x$. The $t+1$ factor is needed to take into account the
fact that the position of the largest number can be anywhere in time.
The index $n$ indicates that the distributions describe the average of
$n$ independent, uniformly distributed random numbers. 

For $t\gg 1$ we have $P_n^t(x) \ll 1$ for most values of $x$, except
for a $1/n$ neighborhood of $1$. 

So
in the integration the most important contribution comes from the part
that is close to $1$. This permits us to restrict the integral to the
part that we can calculate without loss of consistency:
\be\label{Eq_Bn}
\Bn\simeq\int_{1-1/n}^1 (t+1)xp_n(x)P_n^t(x)\d x \quad\quad (t\gg1)
\ee
The probability distribution close to the limits takes the following
forms
\ba
p_n(x)_{x<1/n}&=&{n^n\over (n-1)!}~ x^{n-1}\\
p_n(x)|_{x>1-1/n}&=&{n^n\over (n-1)!}(1-x)^{n-1}
\ea
The cumulative distribution is the integral of the above:
\be
P_n(x)|_{x>1-1/n}=1-{n^{n-1}\over (n-1)!}(1-x)^n
\ee

Let us now turn back to Eq.~\ref{Eq_Bn}. Using an $x=1-y$ variable
replacement and doing integration in parts we arrive at the following
formula after neglecting the exponentially decaying parts:
\ba
\Bn&=&(t+1){n^n\over (n-1)!}\int_0^{1/n}(1-y)y^{n-1}\times
\nonumber\\
&&\times\left(1-{n^{n-1}\over (n-1)!}y^n\right)^t \d y=\nonumber\\
&=&1- \int_0^{1/n}
\left(1-{n^{n-1}\over (n-1)!}y^n\right)^{t+1}\d y
\ea

\bfg
\centerline{\epsfig{file=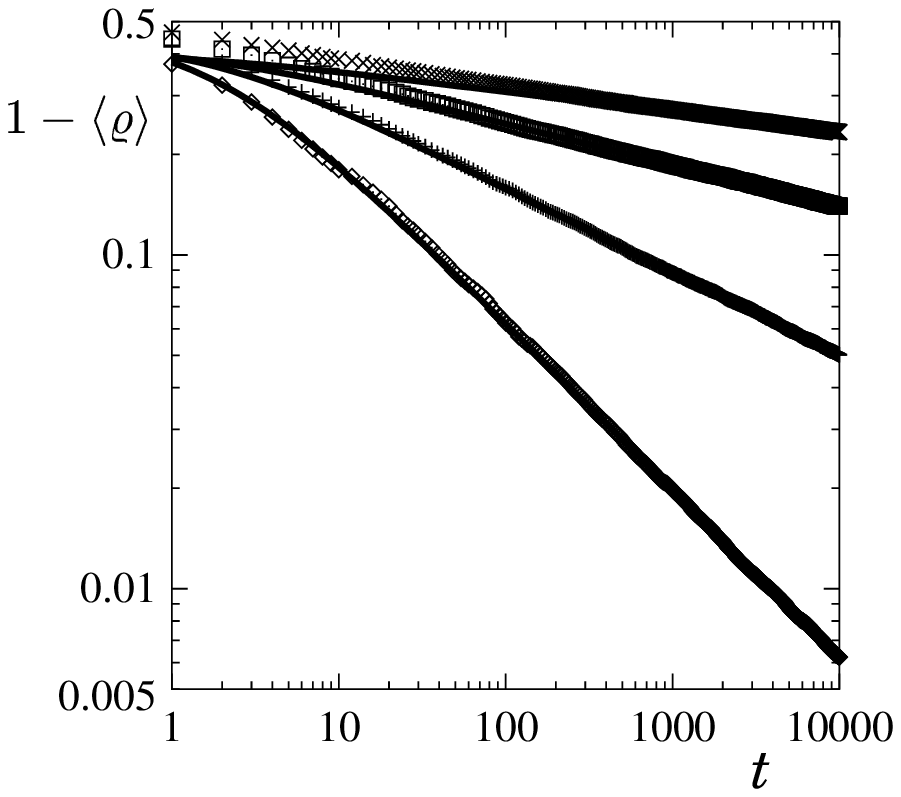,height=6truecm}}
\medskip
\caption{\small \label{Fig_twositeBak} The test of the results in
Eq.~\ref{Eq_twositeresult} against numerical simulations on the same
model. The solid lines are the analytical solutions, the symbols
indicate numerical simulation results. The system sizes are
$n=2(\diamond),\ 4(+),\ 8(\square),\ 16(\times)$ }
\efg

We rewrite the integrand in a $(\cdots)^{t+1}\equiv
\exp\{(t+1)\log(\cdots)\}$ form and make a Taylor expansion in $y^n$
around $y=0$ to the second order. The result can be written in the
following form that we use in our calculations:
\be\label{Eq_twositeresult}
\br{ll}
\Bn=&\displaystyle
1-\G(1/n)(n!)^{1/n} n^{-2}\left(t+{3n+1\over 2n}\right)^{-1/n}+\\
&\displaystyle + {\cal O} \left({1\over t^{1/n+1}}\right)
\sim {1\over e}~t^{-1/n}
\er \ee
On Fig.~\ref{Fig_twositeBak} we can see that the above approximation is
excellent for small $n$ and large $t$.

\end{document}